\documentclass[prl,twocolumn,showpacs,preprintnumbers,amsmathamssymb]{revtex4-1} 
\usepackage{graphicx}
\usepackage{natbib}

\begin{document}

\title{Practical Limitations on Astrophysical Observations of Methanol to Investigate Variations in the Proton-to-Electron Mass Ratio}
\author{Simon Ellingsen$^{1}$}
\author{Maxim Voronkov$^{2}$}
\author{Shari Breen$^{2}$}\affiliation{$^{1}$School of Mathematics and Physics, Private Bag 37, University of Tasmania, TAS 7001, Australia}
\email{Simon.Ellingsen@utas.edu.au}
\affiliation{$^{2}$CSIRO Astronomy and Space Science, PO Box 76, Epping NSW 1710, Australia}

\date{\today}

\begin{abstract}
The possibility of using astrophysical observations of rotational transitions in the methanol molecule to measure, or constrain temporal and spatial variations in the proton-to-electron mass ratio ($\mu$) has recently been investigated by several groups.  Here we outline some of the practical considerations of making such observations, including both the instrumental and astrophysical limitations which exist at present.  This leads us to conclude that such observations are unlikely to be able to improve evidence either for, or against the presence of variations in the proton-to-electron mass ratio by more than an order of magnitude beyond current limits.
\end{abstract}

\pacs{06.20.Jr, 33.15.-e, 98.80.-k, 98.38.Er, 98.58.Ec}

\maketitle



The possibility of using astrophysical observations of rotational transitions in the methanol molecule to measure, or constrain temporal and spatial variations in the proton-to-electron mass ratio ($\mu$) has recently been suggested by several groups \cite{Jansen+11,Levshakov+11}.  The first attempts to use astronomical observations of methanol masers to constrain $\mu$ within the Milky Way show that $|\Delta \mu/\mu| < 28 \times 10^{-9}$ \cite{Levshakov+11}. 

The hindered internal rotation exhibited by methanol produces the degeneracy which makes the different rotational transitions particularly sensitive to the proton-to-electron mass ratio \cite{Jansen+11}.  It also leads to the rich rotational and vibrational spectrum, which due to quantum mechanical selection rules produces a large number of centimetre and millimetre maser transitions in interstellar space.  The strongest and most common of these is the $5_{1} \rightarrow 6_{0} \mbox{A}^{+}$ transition which has a rest frequency of approximately 6.7 GHz.  It has been detected towards more than 900 sites of high-mass star formation in the Milky Way \cite[see for example][]{Caswell+10,Green+10}.  The second strongest astrophysical transition of methanol is from the $2_{0} \rightarrow 3_{-1} \mbox{E}$ transition which has a rest frequency of approximately 12.2 GHz, and is observed towards 43\% of those sources showing 6.7 GHz emission \cite{Breen+11}.  In most methanol maser regions multiple spectral features are observed in the spectra of 6.7 or 12.2 GHz methanol masers at different Doppler shifted velocities \cite{Caswell+10,Green+10}.  The spectral features typically have near-Gaussian profiles (although spectral blending between spatially unresolved components affects most observations), and FWHM (full-width half maximum) of a few tenths of kilometers per second.

The $K_{\mu}$ coefficient measures the sensitivity of a transition to the proton-to-electron mass ratio and for the 6.7 and 12.2 GHz methanol transitions these have been calculated to be -42 and -33 respectively \cite{Jansen+11}.  To improve constraints on $|\Delta \mu/\mu|$, beyond those already achieved \cite{Levshakov+11}, requires measurement of the relative Doppler shifted velocities of the different transitions to an accuracy of better than 100 ms$^{-1}$. When the 6.7 and 12.2 GHz methanol transitions exhibit peaks at the same velocity they have been demonstrated to arise from the same locations at the milliarcsecond level (corresponding to linear scales of a few AU at distances of a few kpc) \cite{Menten+92,Norris+98}.  The large number of 6.7 and 12.2 GHz methanol masers observed within the Milky Way makes them potentially useful for probing spatial variations in $\mu$.  Perhaps more significantly, emission from these two transitions in external galaxies  may allow temporal/cosmological evolution to be investigated.   However, there are two practical issues which need to be considered: 1. The degree to which the emission from the two transitions is coincident/cospatial and 2. The likelihood of being able to detect emission from these transitions at cosmological distances.  We address each of these issues in turn below.



The requirement for using astrophysical observations of different methanol transitions to measure spatial or temporal changes in $\mu$ is that you can measure the velocity (frequency) offset for the two transitions, compared to the laboratory values, and that the observed shift is due to a change in $\mu$.  Levshakov et al. \cite{Levshakov+11} identified that intrinsically strong and narrow spectral lines (e.g. interstellar masers) provide the best opportunity to achieve this, but suggested uncertainties in the rest frequencies of the transitions as the major source of error.  We first review the extensive observational information on methanol masers which has been gathered, primarily over the last 20 years, to determine if there are other possible causes for differences between the observed peak velocity of methanol maser transitions from the same region, and their likely magnitude compared to differences caused by variations in $\mu$.

The 6.7 and 12.2 GHz transitions of methanol are members of the $A$ and $E$ rotational species respectively.  Although chemically identical, they each have independent rotational spectra and are expected to have comparable abundances in interstellar molecular clouds, the most extreme difference in abundance is expected to be approximately 40\% \cite{Sobolev+97}.  At high angular resolution the 6.7 and 12.2 GHz methanol masers are observed to have a complex spatial morphology, with emission present on multiple scale sizes \cite{Minier+02,Harvey-Smith+06}.  The maser emission is typically observed as a series of ``spots'' with bright cores of a few-10 AU, surrounded by weaker emission on scales of 10s-100s of AU \cite{Minier+02}, but sometimes extending to 1000s of AU \cite{Harvey-Smith+06}.  The maser spots usually arise in clusters with linear scales of 6000 AU \cite{Caswell97}.  There are relatively few cases where the spatial structure of individual maser spots has been investigated in different coincident transitions towards the same source, however, where they have it has been found that the structure as measured by plots of the visibility versus baseline length differ \cite{Minier+02}.  This indicates that the spatial scales of the emission from the two ``coincident'' transitions differs.  Individual methanol maser spots are also observed to have internal velocity gradients, sometimes in different directions from the larger scale gradients seen between individual spots within a particular region \cite{Moscadelli+03,Dodson+04}.  Very high spatial (few AU) and spectral (20 ms$^{-1}$) resolution observations of the 12.2 GHz methanol masers in W3(OH) measure velocity gradients of 20-300 ms$^{-1}$AU$^{-1}$, and show deviations from a Gaussian profile are typically at levels less than 0.2\% at these spatial resolutions \cite{Moscadelli+03}.

In the presence of turbulence within the masing gas, if the physical conditions were homogeneous, and multiple maser transitions were saturated and completely cospatial, then the spectra from the different transitions would be identical except for a single scaling factor.  What is observed though, is that the spectra of 6.7 and 12.2 GHz masers towards the same regions are usually significantly different \cite[see for example][]{Breen+11b}.  While in many cases emission is observed from both transitions for many spectral features, and the peak emission is at the same velocity 80\% of the time, the ratio of the intensity for different spectral features of the two transitions varies greatly (usually by more than an order of magnitude) within a single source \cite{Breen+11b}.  Theoretical models of the methanol maser emission show that the intensity of the different transitions can be very sensitive to small changes in physical parameters such as gas temperature and density \cite{Cragg+05}.  The models show that the strongest and most common methanol maser transitions (i.e. the 6.7 and 12.2 GHz) are strongly inverted over a wider range of physical conditions than others such as the $9_{2} \rightarrow10_{1}\mbox{A}^{+}$ (23.1 GHz),  and $3_{1} \rightarrow4_{0}\mbox{A}^{+}$ (107 GHz) transitions \cite{Cragg+05}.  The absence of transitions such as the 23.1 and 107 GHz masers in the majority of sources shows that the physical conditions in most sources must be in the range which favours 6.7 and 12.2 GHz masers, but not the other transitions.  Breen et al. \cite{Breen+11} show that only 43\% of 6.7 GHz methanol masers have an associated 12.2 GHz maser and that in regions where this transition is seen there is a smaller volume of molecular gas conducive to this transition than for the 6.7 GHz.  

The precise geometry of the masers is not known, however, a picture self-consistent with the available observational evidence is that the maser spots are regions within the turbulent molecular gas, where by chance there is an unusually high fraction of the material with velocity coherence along our particular line of sight \cite[see fig. 7c of ][]{Minier+02}.  Under this model, an observer along a different line-of-sight is likely to also detect maser emission from the same region, but from different parts of the cloud, and the linear size of the maser cluster gives a reasonable estimate of both the size of the maser-conducive region and the maximum path length.  Also, the relative intensity of the 6.7 and 12.2 (or any other pair of transitions) from a particular maser spot will depend critically upon the specific physical conditions in the velocity coherent fractions of the path.  These are likely to vary between the various disjoint regions of the path through the molecular gas which produces the maser.  Breen et al. \cite{Breen+11b} show that the variation in the 6.7 versus 12.2 GHz maser luminosity is much less between maser spots/spectral features within an individual source than it is between different star formation regions.  It is also observed that while the variability of 6.7 GHz methanol masers on timescales of decades frequently causes the velocity of the peak spectral feature to change, this comes about through the changes in the relative intensities of different features which were present over the whole period \cite{Caswell+10,Green+10,Ellingsen07}.  It is also clear that the the peak intensity of the stronger masers usually changes by less than 50\%, which is consistent with the intensity of the maser features being governed primarily by the physical conditions over a region much larger than the scale size of individual maser spots.

Methanol is a diamagnetic molecule, so in the presence of a magnetic field it experiences weak Zeeman splitting.  Vlemmings et al.\cite{Vlemmings+11} measured a mean velocity difference of  0.6~ ms$^{-1}$ between the two circular polarizations in a sample of 44 6.7 GHz methanol masers.  So the effects of the magnetic fields in the star formation region on the line profiles of the different transitions are 1-2 orders of magnitude less than the current best estimates of the relative velocity differences, and hence are not likely to add significantly to the measurement uncertainty at present.

\begin{figure*}
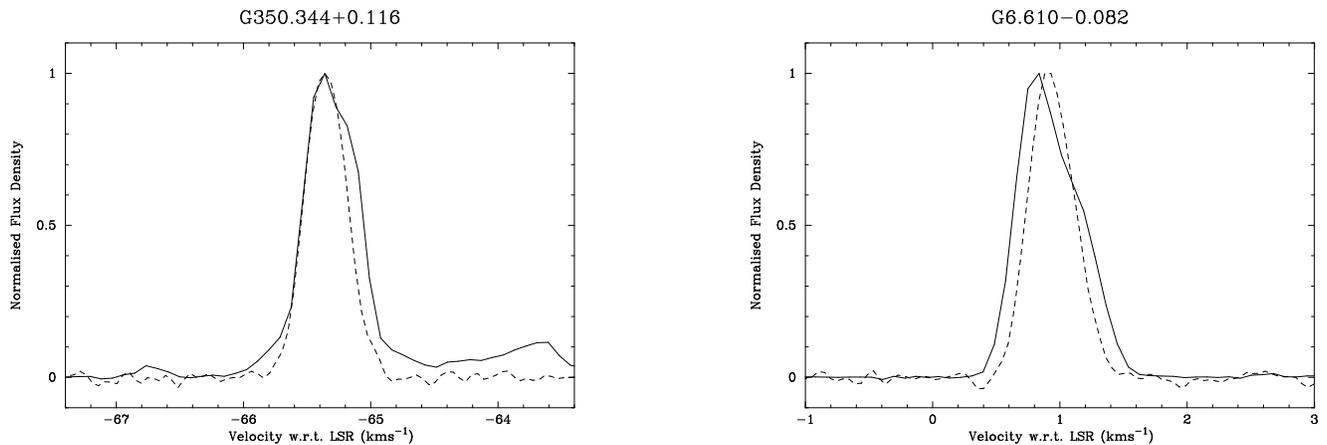

\begin{center}
   \begin{minipage}[t]{0.45\textwidth}
     \includegraphics[angle=270,scale=0.35]{350.344.eps}
   \end{minipage}
   \hfill
   \begin{minipage}[t]{0.45\textwidth}
     \includegraphics[angle=270,scale=0.35]{006.610.eps}
   \end{minipage}
 \end{center}
 \caption{ \label{fig:compare} Comparison of the normalised 6.7 GHz (solid) and 12.2 GHz (dashed) emission in the vicinity of the 12.2 GHz peak for two sources.  The 12.2 GHz spectra are from \cite{Breen+11b}, the 6.7 GHz spectrum of G350.344+0.116 is from \cite{Caswell+10}, while for G6.610-0.082 it is from \cite{Green+10}.}
 \end{figure*}

In practice measuring the relative peak velocities of different interstellar methanol maser transitions to an accuracy greater than 100 ms$^{-1}$ is difficult.  Both high spectral resolution and high sensitivity are necessary, but they alone are not sufficient conditions to achieve the desired accuracy.  Figure~\ref{fig:compare} uses astronomical observations of 6.7 and 12.2 GHz methanol masers to illustrate some of these issues.  We examined the 181 12.2 GHz maser spectra presented by \cite{Breen+11b} and selected 9 sources which showed simple 12.2 GHz spectral profiles with a signal to noise ratio greater than 20.  We then compared the 6.7 GHz spectra from the methanol multibeam observations \cite{Caswell+10,Green+10}, with the 12.2 GHz spectra, after normalising both to the intensity of the peak emission.  The assumed rest frequencies for the two transitions were 6.6685192~GHz and 12.178597~GHz respectively. The spectrum of G350.344+0.116 (see Fig.~\ref{fig:compare}) shows the best match between the normalised 6.7 and 12.2 GHz emission that we found, while the spectrum of G6.610-0.082 shows a second, more typical case.  We fitted one or more Gaussian profiles to the maser spectra and measured a velocity difference between the two transitions of $8 \pm 85$ms$^{-1}$ from a sample of eleven 12.2 GHz spectral peaks (two of the 9 selected sources required two Gaussians to adequately fit the spectrum).  This corresponds to $\Delta \mu/\mu = -2.4 \times 10^{-9} \pm 2.7 \times 10^{-8}$, or an upper limit $|\Delta \mu/\mu| < 27 \times 10^{-9}$, essentially identical to the limit obtained towards a single source by Leshakov et al. \cite{Levshakov+11}.  The observed velocity differences are evenly distributed about the mean (median -2 ms$^{-1}$).  The gas number density ($n_{H_{2}}$) in the regions where the methanol masers arise are around 10$^{6}$ - 10$^{7}$ cm$^{-3}$, hence the limitations on chameleon-like scalar fields implied by these observations are comparable to those obtained from previous observations of NH$_{3}$ and HC$_{3}$N \cite{Levshakov+10}.

We investigated the accuracy to which the central velocity of a maser line can be determined through Monte Carlo simulations of Gaussian profiles observed with similar spectral resolution and signal to noise ratio to our observations.  These simulations suggest that it should be possible to measure the relative velocity of the peaks in the two transitions with an uncertainty of around 8 ms$^{-1}$, approximately an order of magnitude lower than our measured uncertainty.  Figure~4 of Moscadelli et al. \cite{Moscadelli+03} shows that spectra observed on spatial scales greater than a few AU are likely to deviate from Gaussian profiles due to internal velocity gradients. Following \cite{Moscadelli+03} we used the procedure of Watson et al. \cite{Watson+02} to measure the deviation of the selected 12.2 GHz maser spectral profiles from a Gaussian.  We measured $0.003 \pm 0.003$, consistent with a pure Gaussian profile observed with a signal to noise ratio of 20-30, and only marginally higher than that observed with AU-scale spatial resolution \cite{Moscadelli+03}.   However, for each of the selected sources the 6.7 GHz spectra are significantly more complex than the 12.2 GHz, requiring on average five Gaussian components to adequately fit the profiles.  It appears that spectral blending of multiple maser components at the spatial resolution of our observations is the limiting factor in the accuracy of measuring the relative velocities of the 6.7 and 12.2 GHz masers.  Observations of both the 6.7 and 12.2 GHz masers at very high angular and spectral resolution (similar to those of \cite{Moscadelli+03}) are likely to be able to measure the relative velocity of these two transitions with an accuracy of approximately 10 ms$^{-1}$, an order of magnitude better than has been achieved to date.



The primary aim in using astrophysical observations to search for changes in $\mu$ is to make investigations on cosmological timescales, which requires the detection of methanol emission or absorption in distant sources.  Megamasers are maser emission from other galaxies with isotropic luminosities approximately 1 million times greater than typical masers observed in star formation regions in the Milky Way.  They are most commonly observed in the 1667 MHz OH and 22 GHz water transitions and are a distinct class of astrophysical maser, rather than being scaled up versions of the masers seen in Galactic star formation regions.  OH megamasers are observed towards the central regions of LIRGs (Lumninous InfraRed Galaxies) \cite{Darling+02}, while water megamasers are observed toward accretion disks and jets in some active galaxies \cite{Lo05}.  Both OH and water megamasers have been detected at cosmological distances \cite{Darling+02,Lo05}.  

A number of searches for megamaser emission from the 6.7 GHz transition of methanol have been undertaken, primarily targeting relatively nearby galaxies with known OH or water maser emission \cite{Ellingsen+94,Phillips+98,Darling+03}.  No emission was detected in any of these searches, and they are sensitive enough and comprehensive enough, that we can be confident that there are no methanol megamasers with intensities comparable to OH or water megamasers, associated with molecule rich galaxies in the local Universe.  From the observations undertaken to date it appears unlikely that there are any 6.7 GHz methanol megamasers.  This does not rule out the possibility of megamaser emission in some other methanol transition \cite{Sobolev93}, however, methanol maser modelling suggests that the 6.7 GHz transition is the most easily inverted \cite{Cragg+05} and for OH and water, it is the equivalent (commonly detected, easily inverted) transitions which are observed as megamasers.  

A second possibility for detecting methanol masers in external galaxies is through observations which are sufficiently sensitive to detect emission from individual star formation regions.  To date 6.7 GHz methanol masers have been detected in two local group galaxies, the Large Magellanic Cloud (LMC) and Andromeda (M31) \cite{Green+08,Sjouwerman+10}, while sensitive searches in the Small Magellanic Cloud (SMC) and M33 have failed to detect any emission in those Galaxies \cite{Green+08,Goldsmith+08}.  A search of the entire LMC and SMC detected four 6.7 GHz methanol masers, all in the LMC \cite{Green+08}.  Using the observed luminosity distribution of 6.7 GHz methanol masers in the Milky Way, scaled to the distance and star formation rate of the LMC, Green et al. \cite{Green+08} find that the LMC is underabundant in methanol masers by a factor of $\sim$ 5, but has a comparable abundance observed for OH and water masers.  Similarly, the non-detection of 6.7 GHz methanol masers towards 14 star formation regions in M33 suggests that (considering their relative star formation rates), this galaxy is also underabundant in methanol masers compared to the Milky Way \cite{Goldsmith+08}.  It appears that the abundance (and strength) of methanol masers depends strongly on the metalicity of the galaxy \cite{Green+08}, perhaps due to both the presence of two atoms heavier than helium in a methanol molecule, and the higher overall UV field in such galaxies producing more rapid destruction of methanol molecules.  To date there has only been one detection of a 12.2 GHz methanol maser outside the Milky Way, towards the N105a star formation region in the LMC \cite{Ellingsen+10}.  Although it isn't possible to draw strong conclusions from the observations to date, several lines of evidence suggest that the 12.2 GHz masers may be even more sensitive to metallicity than the 6.7 GHz transition \cite{Ellingsen+10,Breen+11}.

With a peak flux density of approximately 5200 Jy \cite{Green+10}, at a distance of 5.2 kpc \cite{Sanna+09}, G9.62+0.20 is one of the most luminous 6.7 GHz methanol masers in the Milky Way.  Were there a 6.7 GHz with similar peak luminosity in the LMC (distance 50 kpc) it would have a peak flux density of around 50 Jy, while in M31 (distance 800 kpc) it would have a peak flux density of around 0.2 Jy.  The strongest 6.7 GHz methanol masers detected in these galaxies have peak flux densities of 4 and 0.012 Jy respectively \cite{Green+08,Sjouwerman+10}, i.e. they are in each case more than an order of magnitude lower luminosity than the strongest Milky Way masers.  There are at least twenty 6.7 GHz methanol masers in the Milky Way with peak luminosity comparable to, or stronger than the strongest known extragalactic masers.  So we consider it likely that most Galaxies will have one or more 6.7 GHz methanol masers with peak luminosity comparable to, or stronger than those seen in the LMC and M31.  Considering future prospects for observing star formation methanol masers in external galaxies, if we assume a receiver performance twice as good as that achieved in the Parkes methanol multibeam survey and an instrument with a square kilometre of collecting area (i.e. the Square Kilometre Array), then in a 1 hour observation (onsource) it would be possible to make a 5$\sigma$ detection of a 6.7 GHz methanol maser with peak luminosity comparable to the strongest detected in the LMC and M31 only out to a distance of approximately 7.5 Mpc (i.e. $z \ll$ 0.01).  

Thermal methanol emission can potentially exhibit much greater line widths than methanol masers, however, it is unlikely that any thermal transitions will have integrated luminosities which exceed the integrated luminosities of the strongest Milky Way methanol masers.  Emission from a number of millimeter thermal methanol transitions (primarily the $2_{k} \rightarrow 1_{k}\mbox{E}$ series at 96.7 GHz), has been observed towards a handful of nearby Galaxies - NGC253, IC342, Maffei 2, NGC6946, NGC4945 and M82 \cite{Henkel+87,Henkel+90,Huttemeister+97,Martin+06}.  The emission from these thermal lines is weak, and broad and offers very little prospect for measuring the small relative changes in rest frequency necessary to constrain or measure changes in $\mu$.  

The most distant detection of a methanol transition is the recent detection of absorption in the $1_{0} \rightarrow 2_{-1}\mbox{E}$ transition (rest frequency 60.5 GHz) towards a $z=0.89$ galaxy in the gravitational lens system PKS\,1830-211 \cite{Muller+11}.  The 12.2 GHz methanol transition is in the same transition family as the 60.5 GHz transition and is frequently observed in absorption towards cold molecular gas within the Milky Way.  For the $z=0.89$ lensing galaxy towards PKS\,1830-211 the absorption from the 12.2 GHz transition lies within the frequency range of both the Very Large Array and the Australia Telescope Compact Array (6.4~GHz).  Measurement of the velocity difference between these two transitions to an accuracy of better than 10~kms$^{-1}$ corresponds to sensitivity to variations in $\Delta \mu/\mu$ at a level of approximately 10$^{-6}$.  Such observations should easily achieve comparable accuracy to the best limits made using ammonia \cite{Murphy+08}, which are comparable to the limits derived from molecular hydrogen \cite{King+08,Thompson+09,Malec+10}.  The intrinsically small size of the background source means a narrow line of sight through the foreground molecular gas, however, changes in the morphology of the continuum emission with frequency will mean different lines of sight for different transitions and provide an additional source of uncertainty.  


In summary, to better constrain variations in $\mu$ with density through observations of Galactic methanol masers (which in turn tests predictions of chameleon-like scalar field theories), requires measurements of the velocity of different methanol maser features accurate to 10s of metres per second, or better.  The comparison of 6.7 and 12.2 GHz maser spectra in a small sample of ``good'' cases shows the observed spectral profiles for the different transitions observed at spatial resolutions greater than a few AU are sufficiently different to produce shifts in the peak velocity of different maser transitions of the order of 100 ms$^{-1}$ in an individual source.  Future observations at high spatial and spectral resolution of both transitions can potentially produce measurements approximately an order of magnitude more accurate than those made to date.  In extragalactic sources sensitive absorption studies would appear to offer the best prospects for extragalactic methanol measurements.  These may enable measurements up to an order of magnitude more sensitive than the best current limits, but have sources of uncertainty in addition to those outlined for Galactic methanol masers.

\end{document}